\documentstyle[12pt,aasms4]{article}
\newcommand{\postscript}[2]
 {\setlength{\epsfxsize}{#2\hsize}
   \centerline{\epsfbox{#1}}}

\begin{document}

\title{Astrometric Properties of Gravitational Binary-Microlens Events\\
       and Their Applications}
\bigskip
\author{Cheongho Han}
\affil{Department of Astronomy \& Space Science, \\
       Chungbuk National University, Chongju, Korea 361-763 \\
       cheongho@astronomy.chungbuk.ac.kr,}
\authoremail{cheongho@astronomy.chungbuk.ac.kr}
\bigskip
\bigskip
\author{Mun-Suk Chun}
\affil{Department of Astronomy, \\
       Yonsei University, Seoul, Korea 120-749 \\
       mschun@galaxy.yonsei.ac.kr,}
\authoremail{mschun@galaxy.yonsei.ac.kr}
\bigskip
\centerline{\&}
\author{Kyongae Chang}
\affil{Department of Physics, \\
       Chongju University, Chongju, Korea 360-764 \\
       kchang@alpha94.chongju.ac.kr}
\authoremail{kchang@alpha94.chongju.ac.kr}

\begin{abstract}
In this paper, we study the astrometric properties of gravitational
microlensing events caused by binary lenses.  By investigating the centroid
shifts for various types of binary-lens events, we find that the deviations
of the centroid shift trajectories from the elliptical ones of
single-lens events are characterized by distortions, twistings, and big
jumps.  We study the conditions of binary-lens system configurations and
source star trajectories for individual types of deviations.  We find
dramatic differences in the astrometric centroid shifts for binary-lens
microlensing events that would be degenerate had their parameters been
determined photometrically.  Therefore, when additional astrometric
observations of a binary-lens event are available, one can resolve the
ambiguity of the binary-lens fit, and uniquely determine the binary-lens
parameters.
\end{abstract}

\vskip20mm

\centerline{submitted to {\it The Astrophysical Journal}: Apr 13, 1999}
\centerline{Preprint: CNU-A\&SS-04/99}
\clearpage


\section{Introduction}
Searches for gravitational microlensing events that monitor millions of
stars located in the Galactic bulge and Large or Small Magellanic Clouds
are being carried out by several groups (MACHO: Alcock et al.\ 1997; EROS:
Ansari et al.\ 1996; OGLE: Udalski et al.\ 1997; DUO: Alard \& Guibert
1997).\markcite{alcock97, ansari96, udalski97, alard97}  Through their 
efforts more than 300 candidate events have been detected.

The light curve of a single-lens event is represented by
$$
A = {u^2+2 \over u(u^2+4)^{1/2}};\qquad
u = \left[\left({t-t_0\over t_{\rm E}}\right)^2+\beta^2\right]^{1/2},
\eqno(1.1)
$$
where $u$ is the lens-source separation in units of the Einstein ring
radius $r_{\rm E}$, $t_{\rm E}$ is the Einstein ring radius crossing time
(the Einstein time scale), $\beta$ is the impact parameter, and $t_0$ is the
time of maximum amplification.  The lensing parameters $(t_{\rm E}, t_0,
\beta)$ of the event are obtained by fitting the observed light curve to
the theoretical curves given by equation (1.1).  Among these parameters, the
Einstein time scale provides information about the lens because it is
related to the physical lens parameters by
$$
t_{\rm E} = {r_{\rm E}\over v};\qquad
r_{\rm E} = \left( {4GM\over c^2}{D_{ol}D_{ls}\over D_{os}}\right)^{1/2},
\eqno(1.2)
$$
where $v$ is the lens-source transverse motion, $M$ is the lens mass, 
and $D_{ol}$, $D_{ls}$, and $D_{os}$ are the separations between the 
observer, lens, and source star.

On the other hand, when the lens is binary, the light curve deviates
from that of the single-lens event in equation (1.1).  The most distinctive
feature of binary-lens event light curve occurs as a source star crosses a lens
caustic (see \S\ 2).  Whenever a source star crosses a caustic, an extra
pair of micro images appears (or disappears), producing a sharp spike in
the light curve, giving rise to ``strong'' binary-lens events.  The optimal
condition for a strong binary-lens event is that the separation between the
two binary components is comparable to the angular Einstein ring radius
$\theta_{\rm E}=r_{\rm E}/D_{ol}$, corresponding to the combined mass of
the binary (combined angular Einstein ring radius).  On the other hand,
when the binary separation is too small or too large, the event has little
chance to involve caustic crossings, resulting in a light curve with a
relatively small deviation from that of a single-lens event (i.e.\ a 
``weak'' binary-lens event; Mao \& Paczy\'nski 1991).\markcite{mao91} 
Currently a total of 13 candidate binary-lens events have been reported.  
These include MACHO LMC\#\hskip-1pt 1 (Dominik \& Hirshfeld 1994, 1996; 
Rhie \& Bennett 1996), OGLE\#\hskip-1pt 6 (Mao \& Di Stefano 1995), 
OGLE\#\hskip-1pt 7 (Udalski et al.\ 1994), DUO\#\hskip-1pt 2 (Alard, Mao, 
\& Guibert 1995), MACHO\#\hskip-1pt 9 (Bennett et al.\ 1996), 95-BLG-12, 
96-BLG-3, 97-BLG-1, 97-BLG-28, 97-BLG-41, 98-BLG-12, 98-BLG-42, and 98-SMC-1
(http://darkstar.astro.washington.edu).\markcite{dominik94, dominik96,
rhie96, mao95, udalski94, alard95, bennett96}

Detecting binary-lens events is important due to their astronomical
applications.  First, one can obtain physical information about binaries
such as mass ratios, $q$, and projected separations, $\ell$.  Second, a
strong binary-lens event provides an opportunity to measure how long it
takes for the caustic line to transit the face of the source star.  By using
the measured caustic crossing time along with an independent determination 
of the source star size, one can measure the lens proper motion relative 
to the observer-source line-of-sight, and thus can estimate the location 
of the lens (Gould 1994; Nemiroff \& Wickramasinghe 1994; Witt \& Mao 1994; 
Peng 1997).\markcite{gould94, nemiroff94, witt94, peng 97} Caustic-crossing 
events can also be used to determine the radial surface brightness profiles 
of source stars (Albrow et al.\ 1999a).\markcite{albrow99a} Finally, 
detection of fine structure in the light curve is an excellent method to 
discover extra-solar planetary systems (Mao \& Paczy\'nski 1991; Gould 
\& Loeb 1992; Griest \& Safizadeh 1998).\markcite{mao91, gould92, griest98}

To extract useful information from binary-lens events, it is essential to
precisely determine the binary-lens parameters (see \S\ 2.1).  However, the
surface of $\chi^2$ behaves in a very complicated manner over the 
multi-parameter space (Albrow et al.\ 1999b).\markcite{albrow99b} As a
result, multiple local minima exist, causing ambiguity in the binary-lens
fit.  

Recently, routine astrometric followup observations of microlensing events
with high precision instruments such as the {\it Space Interferometry
Mission} (hereafter SIM, http://sim.jpl.nasa.gov) have been discussed as a
method to measure the distance and mass of MACHOs (H\o\hskip-1pt g,
Novikov, \& Polnarev 1995; Paczy\'nski 1998; Boden, Shao, \& Van Buren
1998; Han \& Chang 1999).\markcite{hog95, paczynski98, boden98, han99a}
When a microlensing event is due to a single-point lens, the observed
source star image is split into two, and the location of the center of
light (centroid) between the two separate images with respect to the
position of the source star traces out an ellipse (Walker 1995; Jeong, Han,
\& Park 1999).\markcite{walker95, jeong99} However, if the lens is
a binary, both the number and locations of the images differ from those
of a single-lens event, resulting in a centroid shift trajectory that
deviates from an ellipse.

In this paper, we study the astrometric properties of gravitational
microlensing events caused by binary lenses.  By investigating the centroid
shifts for various types of binary-lens events, we find that the deviations
of the centroid shift trajectories from the ellipses of single-lens events
are characterized by distortions, twistings, and big jumps.  We study the
conditions of binary-lens system configurations and source star
trajectories for individual types of deviations.  We also find dramatic
differences in the astrometric centroid shifts for binary-lens events that
would be degenerate had their parameters been determined photometrically.
Therefore, when astrometric observations of binary-lens events
are available in addition to a light curve, one can resolve the ambiguity 
of the binary-lens fit, and thus determine the binary-lens parameters
uniquely.

\section{Binary-lens Events}
When lengths are normalized to the combined Einstein ring radius,
the lens equation of a binary-lens event in complex notation is 
given by
$$
\zeta = z + {m_{1} \over \bar{z}_{1}-\bar{z}} 
          + {m_{2} \over \bar{z}_{2}-\bar{z}},
\eqno(2.1)
$$
where $m_1$ and $m_2$ are the mass fractions of individual lenses, $z_1$ 
and $z_2$ are the positions of the lenses, $\zeta = \xi +i\eta$ and 
$z=x+iy$ are the positions of the source and images, and $\bar{z}$ 
denotes the complex conjugate of $z$ (Witt 1990).\markcite{witt90}  
The amplification of each image, $A_i$, is given by the Jacobian of 
the transformation (2.1) evaluated at the image's position, i.e.,
$$
A_i = \left({1\over \vert {\rm det}\ J\vert} \right)_{z=z_i};
\qquad {\rm det}\ J = 1-{\partial\zeta\over\partial\bar{z}}
             {\overline{\partial\zeta}\over\partial\bar{z}}.
\eqno(2.2)
$$
The images and source positions with infinite amplifications,
i.e.\ ${\rm det}\ J=0$, form closed curves, called critical curves 
and caustics, respectively.  The total amplification of a source 
star is given by the sum of the individual amplifications, i.e.\ 
$A=\sum_i A_i$, with $A_i$ given by equation (2.2).  The position 
of the source star image centroid is the amplification-weighted 
average of the positions of individual images, 
i.e.\ $(x_c,y_c)=(\sum_i A_i x_i/A,\sum_i A_i y_i/A)$, each of
which is obtained by solving the lens equation (2.1) numerically.

\subsection{Light Curves}
To characterize a single-band light curve of a binary-lens event, at least
seven parameters are required.  These include the binary separation $\ell$
and the mass ratio $q$, the closest separation between the source and the
center of the mass of the binary system $\beta$ (impact parameter of a
binary-lens event), the time of the closest approach to the center of mass
$t_0$, the angle that the source star trajectory makes with the projected
binary axis $\alpha$ (approaching angle), the combined Einstein ring radius
crossing time scale $t_{\rm E}$, and finally, the baseline source star flux
$F_0$.  In actual experiments in which observations are performed toward
very dense star fields one should include an additional parameter for the
background flux $B$ to account for blending.  

Due to the combination of a large number of binary-lens parameters and the
complicated behavior of the $\chi^2$ surface over parameter space,
fitting the light curve of a binary-lens event is a formidable task (Albrow
et al.\ 1999b).\markcite{albrow99b}  The most serious problem is that 
there can be multiple local minima, but verifying that one of the local 
minima is a global solution is very difficult.  The ambiguity in 
binary-lens fits is demonstrated well by Dominik (1999).\markcite{dominik99} 
He showed that several models having a large variety of parameters agree 
well with the observed binary-lens events OGLE\#\hskip-1pt 7 and 
DUO\#\hskip-1pt 2.  In Figure 1, we reproduce four different model light 
curves based on the solutions to the binary-lens parameters determined 
by Dominik (1999)\markcite{dominik99} and the measured $I$-band data 
points of the binary-lens event OGLE\#\hskip-1pt 7 (Udalski et al.\ 
1994).\markcite{udalski94} In Figure 2, we present the configurations of 
the binary-lens system and the source star trajectories for the individual 
model fits corresponding to the model light curves in Figure 1.  In Figure 
2, the projected locations of the lenses are marked by `x' and the caustics 
and source star trajectories are represented by the thick curves and thin 
straight lines respectively.  In each panel of Figure 2, we also mark 
the values of the binary-lens parameters $\ell$, $q$, and $t_{\rm E}$.  
One finds that despite the similarity of the light curves, the resulting 
binary-lens parameters for individual model fits are significantly 
different from each other.

\subsection{Centroid Shifts of Source Star Images}
When a source star is microlensed by a single point-mass lens, its image 
is split into two.  The typical separation between the two images for a
Galactic bulge event caused by a typical stellar mass lens is on the
order of a milliarcsecond, and thus one cannot resolve the individual images.
However, the source star image centroid shifts caused by microlensing can
be measured with high-precision interferometers such as the SIM that will
have a positional accuracy of $\lesssim 10\ \mu{\rm arcsec}$.  The centroid
shift for a single-lens event is related to the lensing parameters by
$$
\vec{\delta\theta}_{c} = {\theta_{\rm E} \over u^2+2}
\left[ \left( { t-t_0\over t_{\rm E}}\right)\hat{\bf x} + 
\beta\hat{\bf y}\right];\qquad
\theta_{\rm E} = \left( {4GM\over c^2}{D_{ls}\over D_{ol}D_{os}} \right)^{1/2},
\eqno(2.2.1)
$$
where $\theta_{\rm E}$ is the angular Einstein ring radius, 
the $x$ and $y$ axes represent the directions which are parallel 
and normal to the lens-source transverse motion and $\hat{\bf x}$
and $\hat{\bf y}$ are the unit vectors for these corresponding directions.
If we let $x=\delta\theta_{c,x}$ and $y=\delta\theta_{c,y}-b$, where
$b = \beta/2(\beta^2+2)\theta_{\rm E}$, the coordinates are related by
$$
x^2 + {y^2\over {\cal Q}^2} = a^2,
\eqno(2.2.2)
$$
where $a = \theta_{\rm E} / 2(\beta^2+2)^{1/2}$
and ${\cal Q} = b/a = \beta / (\beta^2+2)^{1/2}$.
Therefore, the trajectory of the apparent source star image centroids
traces out an ellipse during the event, known as the `astrometric ellipse'
(Walker 1995; Jeong et al.\ 1999).\markcite{walker95, jeong99}

If a lens is binary, on the other hand, the number and positions of the 
source star images differ from those of a single-lens event(Safizadeh, 
Dalal, \& Griest 1998).\markcite{safizadeh98}  As a result, the centroid 
shift trajectory of the binary lens event deviates from an ellipse.
To see the pattern of the deviations, we compute the centroid shifts 
for events caused by binaries with various separations and axis ratios 
and show their trajectories in Figure 3.  In Figure 4, we also present 
the lens system configurations and source star trajectories which are 
responsible for the centroid shifts in Figure 3.  In Figure 4, the caustics 
and positions of the lenses are represented in the same manner as in 
Figure 2 and the source star trajectories are represented by straight 
lines with line types chosen to match those of the centroid shift 
trajectories in Figure 3.  The impact parameters of the individual source 
star trajectories are $\beta=0.3$ (solid lines), 0.5 (dotted lines), and 
0.7 (dashed lines), and the approaching angle $\alpha$ is $20^\circ$ for 
all trajectories.

From Figure 3, one finds that the diversity of the centroid shift trajectories 
expected from binary-lens events is very large, similar to photometric light 
curves.  However, despite the large variations, one finds that the deviations 
of the centroid shift trajectories from the astrometric ellipses of single-lens
events can be classified by having {\it distortions}, {\it twistings}, and 
{\it big jumps}.  First, distortions in the trajectories of centroid shifts 
occur for non-caustic-crossing binary-lens events in which binary separations 
are relatively small compared to their impact parameters.  For an event with 
a very close binary separation, i.e.\ $\ell\ll\beta$, the centroid shift 
trajectory is well approximated by the astrometric ellipse of a single-lens 
event with mass equal to the total mass of the binary and located at the 
center of mass of the binary.  However, we note that for the same binary-lens 
event the fractional deviation of the astrometric centroid shifts from those 
of the corresponding single-lens event is significantly larger than the 
fractional deviation of the photometric amplifications, making astrometrically 
observing microlensing events a useful method to detect very close binaries 
(Chang \& Han 1999).\markcite{chang99} The amount of deviation increases with 
an increasing ratio of $\ell/\beta$.  Secondly, when the binary separation 
is equivalent to the impact parameter, i.e.\ $\ell\sim\beta$, the trajectory 
of the centroid shift becomes twisted, developing a loop near the closest 
approach.  For a given lens system configuration, the size of the loop is 
bigger for events with smaller impact parameters.  We note that similar 
distortions and twistings occur due to the effect of finite source size (Mao 
\& Witt 1998).\markcite{mao98}  However, while the source star trajectories 
deviated by the finite source effect are symmetric with respect to the line 
connecting the position of the unlensed source star and the centroid at the 
time of maximum amplification, the centroid shift trajectories caused by 
binary lenses are, in general, asymmetric.  Therefore, one can distinguish 
the trajectories of astrometric centroid shifts caused by binary lenses from 
those affected by the finite source effect.  Finally, whenever the source star 
crosses a caustic, the centroid of the source star images shifts by a large 
amount, up to about $0.''\hskip-1pt 001$ - $0.''\hskip-1pt 01$ (Witt \& Mao 
1996),\markcite{witt96} producing a big jump in the centroid shift trajectory.

\subsection{Application of Astrometric Observations}
In the previous subsection, we showed that, depending on the lens-system 
configurations and source star trajectories, i.e.\ binary-lens parameters, 
the trajectories of the centroid shifts of binary-lens events take 
various forms.  Thus, one can determine the binary-lens parameters from 
the centroid shift trajectory in a manner similar to the determination 
of parameters from the light curve.  To fit a centroid shift trajectory, 
however, the number of parameters is almost the same as the number to fit 
a light curve.
\footnote
{
For an actual observation of a microlensing event, there are two more 
parameters needed to describe the centroid shift than are needed to 
describe the light curve.  These parameters are used to represent the 
position of the blended star $(x_b,y_b)$ because the centroid shifts are 
not only affected by the flux of the blended star, but also by its location.  
However, one can correct for the effect of nearly all types of blending, 
and thus both the values of $(x_b,y_b)$ and $B$ can be determined by using 
the diffraction-limited resolution of space-borne observations and the 
high positional accuracy of the SIM (Han \& Kim 1999).\markcite{han99b}  
Therefore, the effective number of parameters necessary to describe the 
centroid shift trajectory is one less than the number of parameters needed 
to characterize the light curve.
}
The result is that the astrometric determination of the binary-lens parameters  
suffers from the same degree of ambiguity as the photometric determination. 
Therefore, there is no advantage to determining the binary-lens parameters 
from the centroid shifts alone.

However, one can completely resolve the ambiguity in finding the parameters
by using both the photometric light curve and the astrometric centroid 
shift trajectory.  This is because the astrometric centroid shifts expected 
from the photometrically determined  binary-lens parameters are significantly 
different from each other.  To demonstrate this, in Figure 5 we present the 
trajectories of the centroid shifts that are expected for the four degenerate 
lens-parameter solutions corresponding to the model light curves in Figure 1.
From Figure 5 one finds that despite the very similar the light curves, 
the trajectories of the centroid shifts differ dramatically from each other.  
To isolate a global solution, the measured centroid shift trajectory is 
simply compared to the solutions producing the photometric fit.

\section{Summary}
The findings from our investigation of the astrometric properties of 
binary-lens gravitational microlensing events are summarized as follows.
\begin{enumerate}
\item
The trajectories of centroid shifts expected from binary-lens events 
take various forms depending on the lens-system configurations and source 
star trajectories.
\item
Despite the large diversity of centroid shift trajectories, the deviations 
from the elliptical trajectories of single-lens events can be categorized 
into distortions, twistings, and large jumps. 
\item
The centroid shift trajectories expected from different degenerate solutions
of the photometrically determined binary-lens parameters are dramatically 
different from each other.  Therefore, with the additional information 
provided by the astrometrically determined centroid shifts, one can completely 
resolve the ambiguity of the photometric binary-lens fit, and thus can 
uniquely determine the binary-lens parameters.
\end{enumerate}

\acknowledgements
We would like to thank M.\ Everett for careful reading of the manuscript.  
M.\ S.\ Chun was supported by the financial grant from the Korea 
Research Foundation (1998-015-D00287) in the program year of 1999.

\clearpage

\postscript{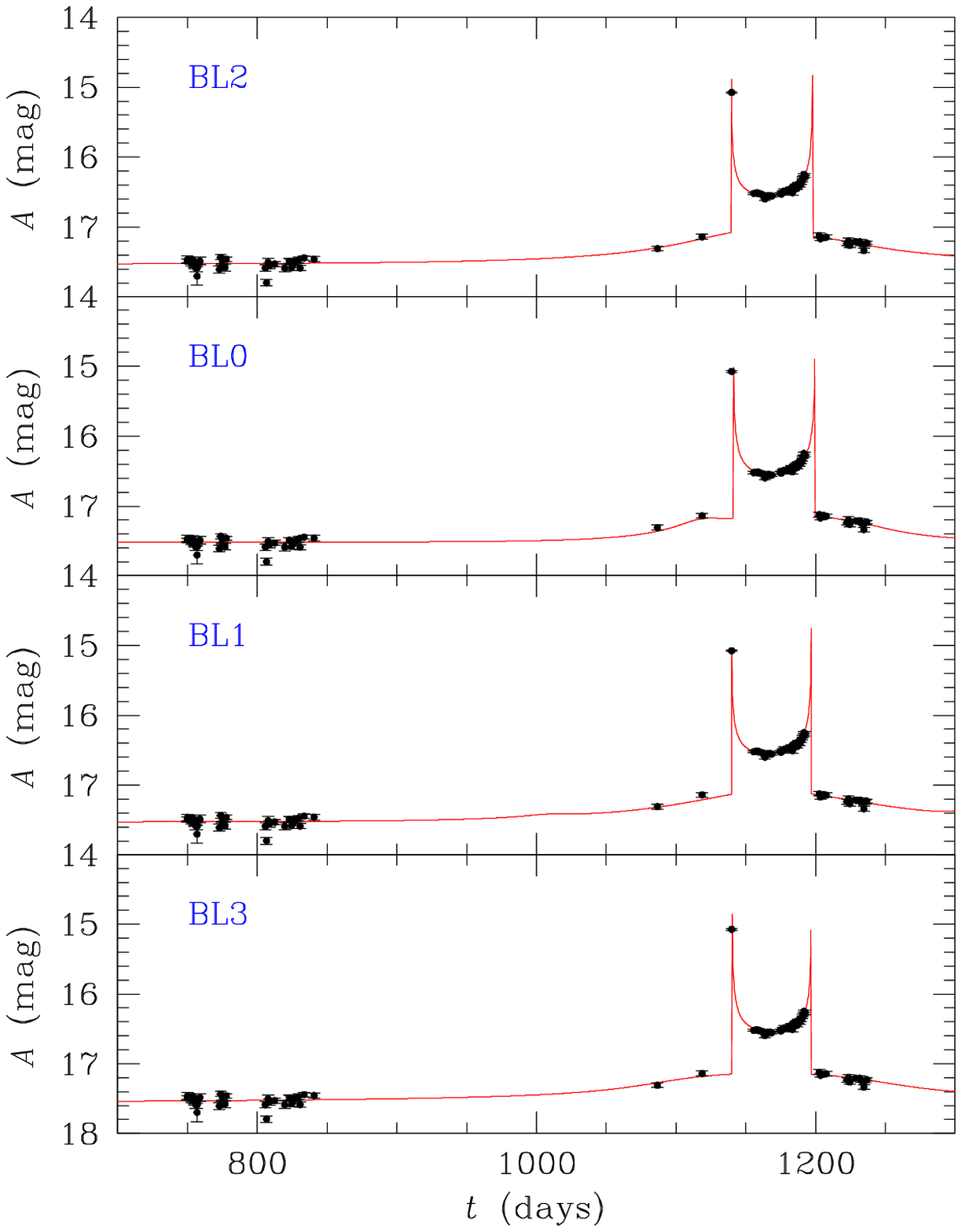}{1.0}
\noindent
{\footnotesize {\bf Figure 1:}\
The measured $I$-band data points of the binary-lens event 
OGLE\#\hskip-1pt 7 and the four different model light curves (solid lines).
The designations of individual model light curves (BL2, BL0, BL1, and BL3) 
are borrowed from Dominik (1999) who originally determined the binary-lens 
parameters for the model light curves.
} \clearpage

\postscript{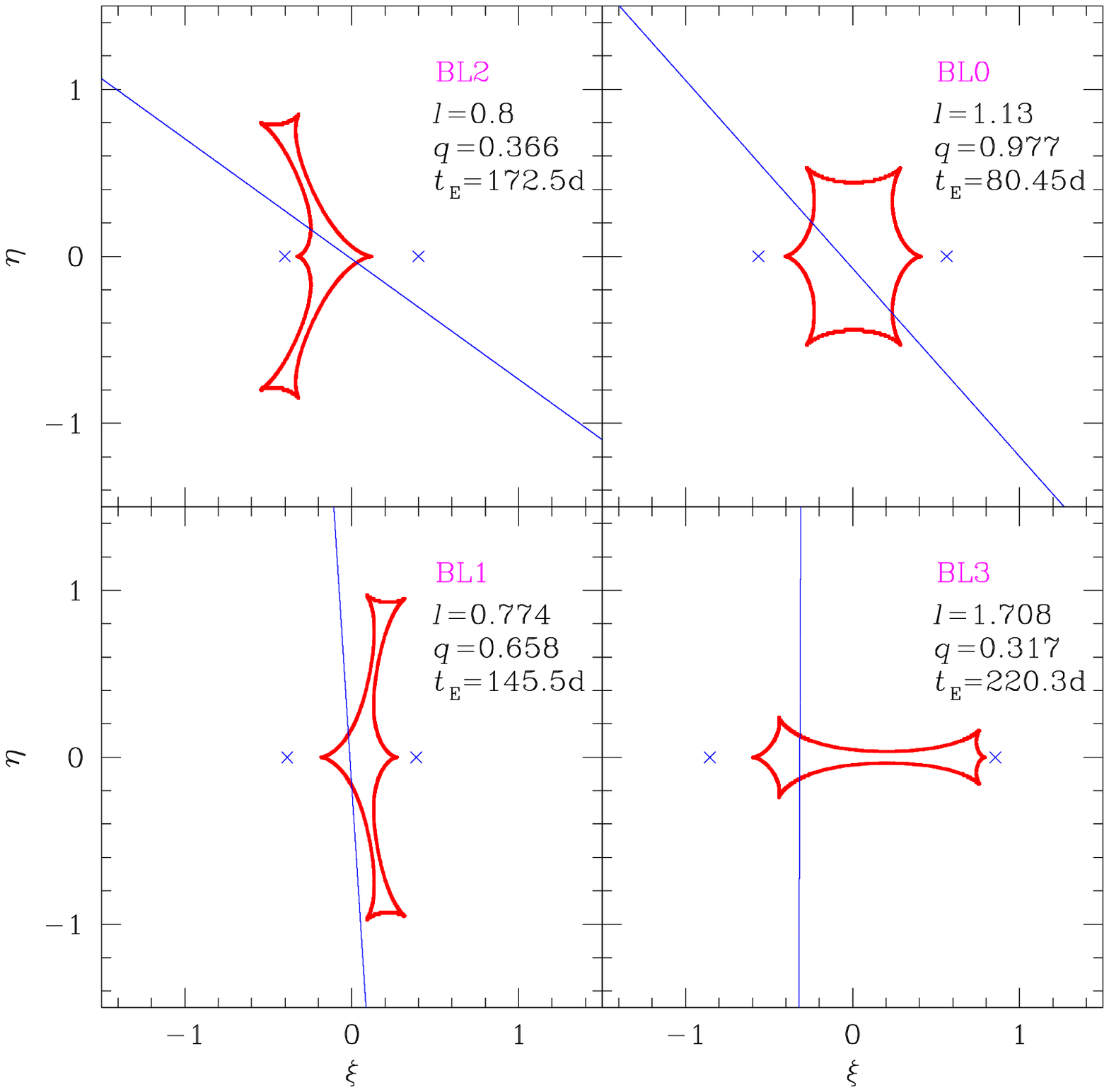}{1.0}
\noindent
{\footnotesize {\bf Figure 2:}\
The configurations of the binary-lens systems and source star trajectories 
(thin solid lines) that correspond to the model light curves in 
Figure 1.  For each configuration, the lens locations and the caustics are 
marked by `x' and thick solid lines respectively.  Also marked in each panel are 
the values of the binary-lens parameters $\ell$, $q$, and $t_{\rm E}$.
One finds that despite the similarity of the light curves (in Figure 1), 
the difference in the binary-lens parameters are significant.
} \clearpage

\postscript{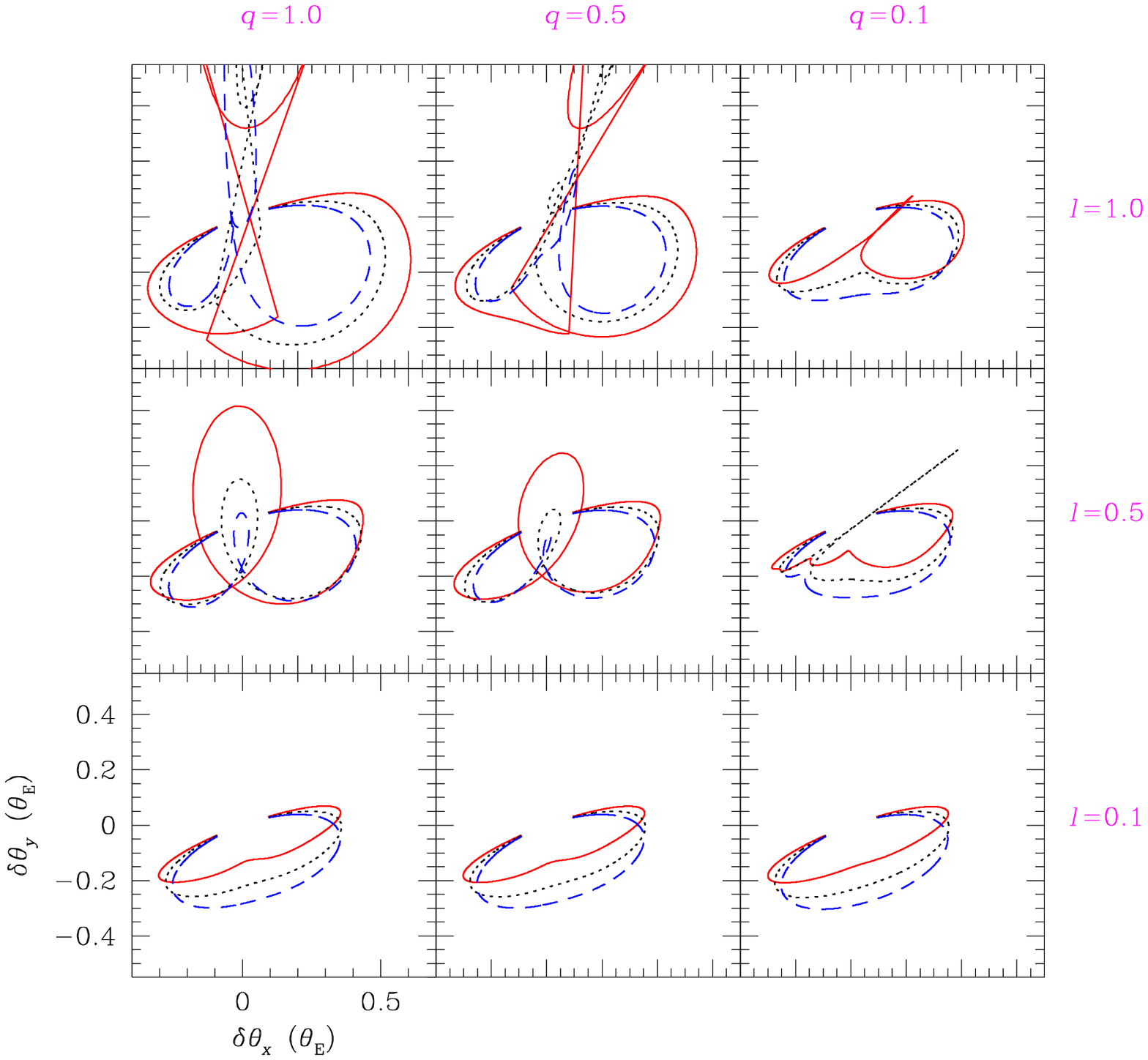}{1.0}
\noindent
{\footnotesize {\bf Figure 3:}\
Trajectories of source star image centroid shifts for microlensing events 
caused by binary lenses with various binary separations, $\ell$, and mass 
ratios $q$.  
} \clearpage

\postscript{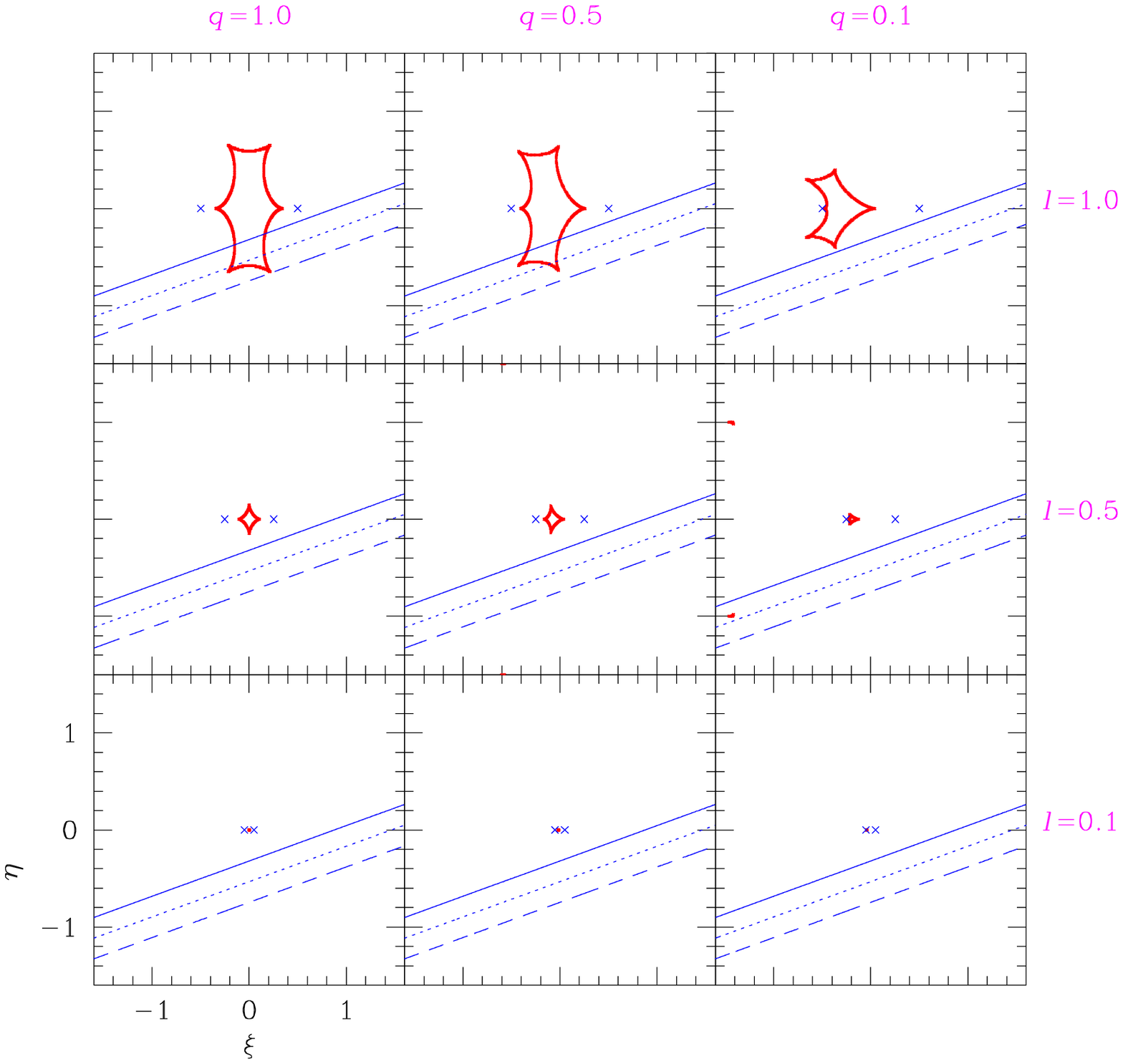}{1.0}
\noindent
{\footnotesize {\bf Figure 4:}\
The lens system configurations and source trajectories (thin solid lines)
which are responsible for the centroid shift trajectories in Figure 3.  
The caustics and the positions of lenses are represented in the same way 
as in Figure 2.  The trajectories have the closest separations of $\beta=0.3$ 
(solid lines), 0.5 (dotted lines), and 0.9 (dashed lines) and have the 
same approaching angle of $\alpha=20^\circ$.  The line type of each source 
star trajectory is chosen so that it matches with the line type of the 
centroid shift trajectory in Figure 3.
} \clearpage

\postscript{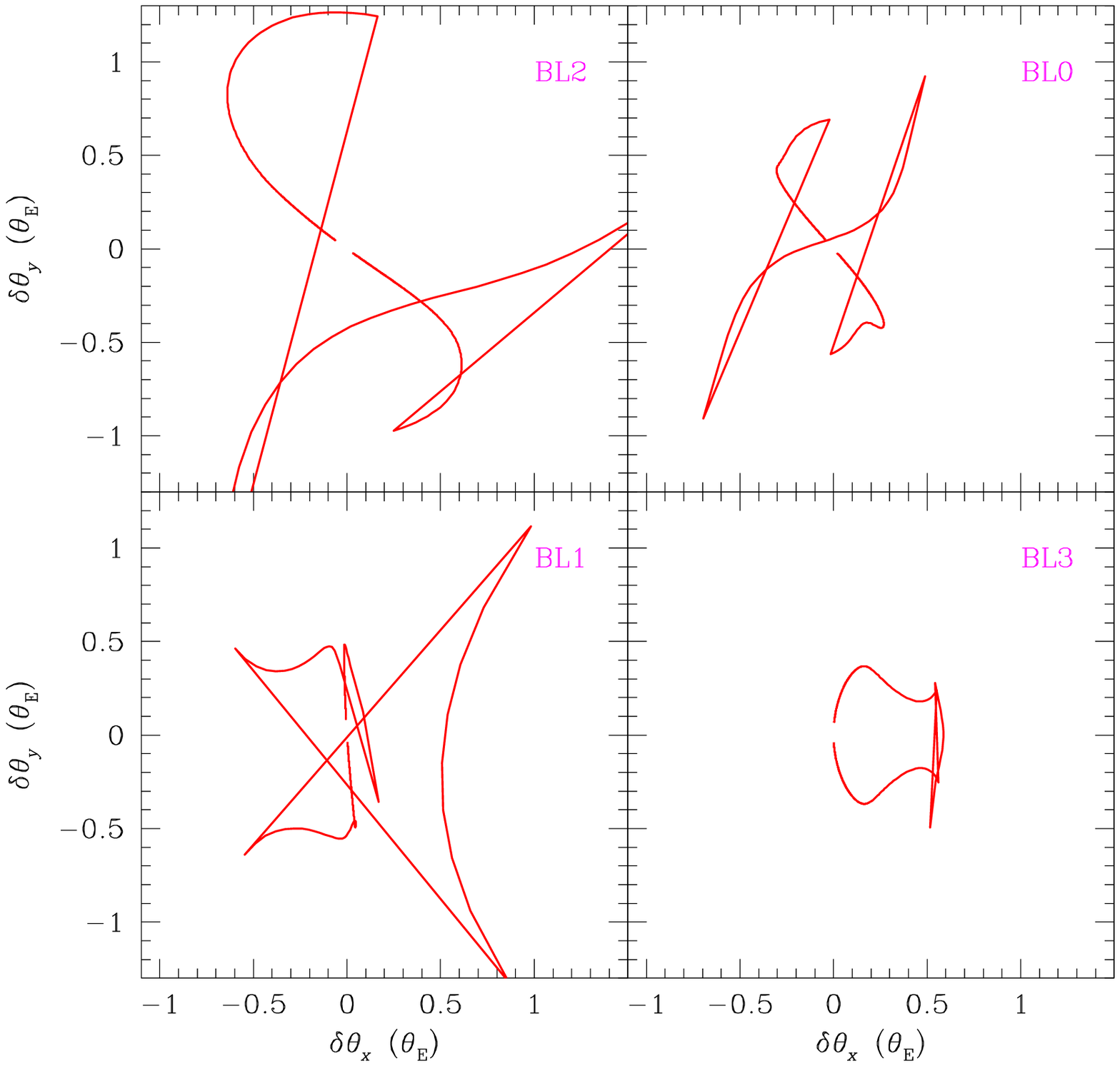}{1.0}
\noindent
{\footnotesize {\bf Figure 5:}\
The centroid shift trajectories that are expected for the four 
lens parameters solutions corresponding to the model light curves 
in Figure 1.  One finds that despite the similarity in the light curves,
the centroid shift trajectories are dramatically different from one another.
} \clearpage

\end{document}